\documentstyle[aps,epsf]{revtex}
\begin{document}

\title{Theory of
solitons, polarons and multipolarons in one dimension:
An alternative formulation}
\author{Kihong Kim\cite{kk}}
\address{Department of Molecular Science and Technology, 
Ajou University, Suwon 442-749, Korea}
\author{Dong-Hun Lee}
\address{Department of Astronomy and Space Science, 
Kyung Hee University, Yongin 449-900, Korea}
\maketitle 
\begin{abstract}
We develop an alternative formulation of 
the theory of solitons, polarons and multipolarons
in quasi-one-dimensional degenerate 
and non-degenerate conducting polymers,
starting from the continuum Hamiltonian 
introduced by Brazovskii and Kirova.
Based on a convenient real-space representation 
of the electron Green function in one dimension, we present
a simple method of calculating the Green function 
and the density of states 
in the presence of a single soliton or polaron defect,
using which we derive exact expressions for the soliton, polaron
and multipolaron excitation energies and the 
self-consistent gap functions  
for an arbitrary value of the electron-phonon coupling constant.
We apply our results to $cis$-polyacetylene.
\\ 
\end{abstract}

\section{Introduction}
\label{sec-int}

A wide variety of fascinating physical phenomena occur in 
quasi-one-dimensional materials with a Peierls-Fr\"olich 
ground state \cite{gru,roth}.
In particular, the proposal \cite{braz,ssh,rice,bk2,cb} that 
non-linear excitations such
as solitons and polarons play a crucial
role in the electronic properties 
of conjugated conducting polymers
and other related materials
have attracted great interest over the past two decades
\cite{roth,bk1,camp,hkss,kiess}.
Experimental and theoretical efforts to confirm the existence of 
these non-linear excitations
and to clarify their properties are
being continued to the present day \cite{eps1,eps2,oka,kw,mos}.

In a large number of theoretical studies
of solitons and polarons, 
the electron-electron interaction
is ignored \cite{int} and the lattice motion is treated classically.
A simple lattice model of non-interacting electrons 
in one dimension coupled 
to phonons was introduced by Su, Schrieffer and Heeger \cite{ssh}
and has been applied mainly to numerical studies of 
$trans$-polyacetylene [(${\rm CH})_x$], 
which is a representative 
conjugated polymer with two degenerate ground states. 
A continuum version of this model
derived by Takayama, Lin-Liu and Maki \cite{tlm} 
admits exact soliton and polaron
solutions 
and has been the starting point
in a number of analytical studies \cite{cb,fbc}.
In order to treat non-degenerate materials 
with a unique ground state and a higher-energy metastable state
such as $cis$-polyacetylene, Brazovskii and Kirova
derived a generalized version of the continuum model,
which contains a new parameter, $\Delta_e$, representing
the {\it extrinsic} gap in the electronic spectrum 
that exists even in the absence
of the Peierls distortion \cite{bk2}.
When $\Delta_e=0$, Brazovskii and Kirova's model reduces to
Takayama, Lin-Liu and Maki's model. If $\Delta_e\ne 0$, 
it does not admit the soliton solution, but allows the so-called
multipolaron solutions in addition to the polaron solution. 

In this paper, we reformulate the theory of solitons, 
polarons and multipolarons,
starting from the continuum Hamiltonian introduced 
by Brazovskii and Kirova, which can describe both 
degenerate and non-degenerate conducting polymers depending 
on the value of $\Delta_e$.
We develop an efficient method for calculating the Green functions
associated with the non-linear excitations,
based on a convenient real-space representation 
of the Green function in one dimension. 
Using the density of states expression
obtained from the Green function,
we compute the soliton, polaron and multipolaron
excitation energies and the related 
self-consistent gap functions
{\it exactly} for an {\it arbitrary} value of the electron-phonon coupling 
constant. These results are usually presented in the weak 
electron-phonon coupling limit in the literature
and, to the best of our knowledge, have 
never been written down explicitly before.
Our method, which involves only concepts 
familiar in condensed matter physics,
is simple and transparent and has the advantage of
being readily generalizable to more difficult problems 
such as the influence of disorder on solitons and polarons.

In the next section, we introduce the Hamiltonian and the
soliton wave function. The polaron case will be discussed in 
Appendix A.
In Sec.~\ref{sec-gf} and Appendices A and B, 
we describe our method for calculating 
the Green function and the density of states.
In Sec.~\ref{sec-res}, we compute the soliton, polaron and 
multipolaron excitation energies and the self-consistent 
gap functions and apply the results to $cis$-polyacetylene. 
In Sec.~\ref{sec-conc}, we conclude the paper with some remarks.

\section{Hamiltonian and wave functions}
\label{sec-model}

We consider the continuum Hamiltonian of 
quasi-one-dimensional conducting polymers
and related materials
first introduced by Brazovskii and Kirova \cite{bk2,cont}:
\begin{eqnarray}
H&=&\sum_{s}\int dx ~\Psi_s^\dagger(x)
\left\{-i\hbar v_F\sigma_3{d\over
{d x}}+\left[\Delta_i(x)
+\Delta_e\right]\sigma_1\right\}\Psi_s(x)
\nonumber\\
&+&{1\over{\pi\hbar v_F\lambda}}\int dx~ \Delta_i^2(x),
\label{eq:bk}
\end{eqnarray}
where $v_F$ is the Fermi velocity and $\lambda$
is the dimensionless electron-phonon coupling
constant (For a precise definition of $\lambda$
in terms of the parameters of the lattice model, 
refer to \cite{boya}). $\sigma_i$'s ($i=1,2,3$)
are Pauli matrices.
The (real-valued) gap function is written as $\Delta(x)=\Delta_i(x)
+\Delta_e$, where 
$\Delta_i(x)$ is the {\it intrinsic} gap function,
which is  
sensitive to electron-phonon coupling and 
varies slowly on the scale of a lattice spacing, and
$\Delta_e$ is a constant {\it extrinsic} component.
The time 
derivative of $\Delta_i(x)$ is assumed to be 
sufficiently small that the lattice
kinetic energy can be ignored.
$\Psi_s(x)$, $s=\uparrow,\downarrow$ being
the spin index, is the two-component electron wave function
for non-interacting electrons moving to the right ($\psi_{1s}$) 
and to the left ($\psi_{2s}$):
\begin{equation}
\Psi_s(x)=\pmatrix{\psi_{1s}(x)\cr \psi_{2s}(x)\cr}.
\end{equation}
From the Hamiltonian, we obtain the Dirac-type equation for
the electron wave functions
\begin{equation}
\left( \begin{array}{cc}
-i\hbar v_F \displaystyle{\frac{d}{dx}}&
\Delta(x)\\
\Delta(x)&
i\hbar v_F \displaystyle{\frac{d}{dx}}
\end{array}\right)        
\left( \begin{array}{c}\psi_{1s}(x)\\ \psi_{2s}(x) \end{array}\right)
=E\left( \begin{array}{c}\psi_{1s}(x)\\ \psi_{2s}(x) \end{array}\right)
\label{eq:main}
\end{equation}
and the self-consistent gap equation
\begin{equation}
\Delta_i(x)=\Delta(x)-\Delta_e=-{{\pi\hbar v_F\lambda}\over 2}
{\sum}^\prime\left[\psi_{1s}(x)\psi_{2s}^*(x)
+\psi_{2s}(x)\psi_{1s}^*(x)\right],
\label{eq:gap}
\end{equation}
where the prime on the summation symbol indicates the sum
is over all {\it occupied} electronic states.

We distinguish between the 
two cases with $\Delta_e=0$ and $\Delta_e>0$.
Let us first consider the case with $\Delta_e=0$.
Then the Hamiltonian has two degenerate
ground states with $\Delta(x)=\Delta_i(x)=\pm \Delta_0$, 
where $\Delta_0$ ($>0$) is 
the self-consistent uniform Peierls gap parameter.
In this paper, we focus on non-uniform
solutions for the gap function. 
The best-known ones are the single soliton and the single polaron.
Here we discuss the soliton and put all the algebra for the polaron
in Appendix A. The spatial variation is set by the coherence length,
a scale defined by the Fermi velocity and the uniform gap:
\begin{equation}
\xi=\frac{\hbar v_F}{\Delta_0}.
\label{eq:cohl}
\end{equation}
A second dimensionless parameter $\kappa$ is important for polarons 
and is defined and discussed in Appendix A.
The spatial variation of the order parameter for a single 
soliton located at $x=0$ is
\begin{equation}
\frac{\Delta(x)}{\Delta_0}=\tanh\left( \frac{x}{\xi} \right).
\label{eq:sol}
\end{equation}
The resulting (unnormalized) wave functions for 
unbound states with $\vert E\vert\ge\Delta_0$
can be written as 
\begin{eqnarray}
\psi_{1s}(x)&=&\phi_1(x)=e^{ikx}\left[\tanh\left(\frac{x}{\xi}\right)
-i\frac{E}{\Delta_0}-ik\xi\right],\nonumber\\ 
\psi_{2s}(x)&=&\phi_2(x)=-ie^{ikx}\left[\tanh\left(\frac{x}{\xi}\right)
+i\frac{E}{\Delta_0}-ik\xi\right],
\label{eq:solwf1}
\end{eqnarray}
where the quantum number $k$ is related to the energy eigenvalue by 
$E^2=\left(\hbar v_F k\right)^2+{\Delta_0}^2$ \cite{cb2}.
For a reason to be explained in the next section, we introduce another
equivalent set of wave functions
\begin{eqnarray}
\psi_{1s}(x)&=&\tilde\phi_1(x)=ie^{-ikx}\left[\tanh\left(\frac{x}{\xi}\right)
-i\frac{E}{\Delta_0}+ik\xi\right],\nonumber\\ 
\psi_{2s}(x)&=&\tilde\phi_2(x)=e^{-ikx}\left[\tanh\left(\frac{x}{\xi}\right)
+i\frac{E}{\Delta_0}+ik\xi\right],
\label{eq:solwf2}
\end{eqnarray}
which are trivially obtained from Eq.~(\ref{eq:solwf1}) by replacing $k$ with $-k$ 
and multiplying the wave functions by $i$.
We omitted the spin indices
in the notations $\phi_1$, $\phi_2$, $\tilde\phi_1$ and $\tilde\phi_2$
because the wave functions are spin-independent. 
There is also a soliton bound state located at $E=0$
and described by the wave functions
\begin{equation}
\psi_{1s}(x)=\frac{1}{2\sqrt{\xi}}{\rm sech}\left(\frac{x}{\xi}\right),~~~
\psi_{2s}(x)=\frac{-i}{2\sqrt{\xi}}{\rm sech}\left(\frac{x}{\xi}\right).
\end{equation}
Though there is
no electronic state for $0<\vert E\vert<\Delta_0$, we can still use 
the wave functions (\ref{eq:solwf1}) and (\ref{eq:solwf2}) for
computing the subgap Green functions.
 
When the extrinsic component of the gap function, $\Delta_e$,
is non-zero, the double-degeneracy of the ground state is broken 
and there appears a higher-energy metastable state with a uniform gap.
For $\Delta_e>0$, the constant intrinsic gap $\Delta_i$ 
can be taken to be positive in the Peierls ground state
and negative in the metastable state \cite{boya}.
In both states, the uniform Peierls gap parameter is given by $\Delta_0
=\vert \Delta_i +\Delta_e\vert$. We caution the readers that we will use 
the same notation $\Delta_0$ and the definition (\ref{eq:cohl}), 
regardless of the value of $\Delta_e$. 
It turns out that the soliton solution is not 
allowed in non-degenerate cases because 
the equations (\ref{eq:sol}) and (\ref{eq:solwf1}) do not satisfy
the gap equation (\ref{eq:gap}) for non-zero $\Delta_e$. 
 
\section{Green function and the density of states}
\label{sec-gf}

In this section, we compute 
the retarded and advanced $2 \times 2$ matrix
Green functions $G^+$ and $G^-$ associated with Eq.~(\ref{eq:main}). 
For that purpose, we introduce two 
linearly-independent wave functions $\psi$ and 
$\tilde \psi$ that satisfy Eq.~(\ref{eq:main}) 
in the interval $-L\le x\le L$ and 
the boundary conditions that $\psi(-L)$ 
and $\tilde\psi(L)$ do not diverge 
as $L$ becomes large, whereas $\psi(L)$ 
and $\tilde\psi(-L)$ diverge \cite{kw}.
We write the Green function in the form
\begin{eqnarray}
&&G(x,y\vert E) = \left( \begin{array}{cc}
G_{11}(x,y \vert E) & G_{12}(x,y\vert E)\\
G_{21}(x,y \vert E) & G_{22}(x,y \vert E) 
\end{array} \right) \nonumber \\
&&~= \left\{ \begin{array}{ll}
\displaystyle{\frac{i}{\hbar v_F(\psi_1\tilde\psi_2-\psi_2\tilde\psi_1)}}
\left( \begin{array}{cc}
\tilde\psi_1(x)\psi_2(y) & \tilde\psi_1(x)\psi_1(y) \\
\tilde\psi_2(x)\psi_2(y) & \tilde\psi_2(x)\psi_1(y) \end{array}\right)
& \mbox{if}~ x>y \\
\displaystyle{\frac{i}{\hbar v_F(\psi_1\tilde\psi_2-\psi_2\tilde\psi_1)}}
\left( \begin{array}{cc}
\psi_1(x)\tilde\psi_2(y) & \psi_1(x)\tilde\psi_1(y) \\
\psi_2(x)\tilde\psi_2(y) & \psi_2(x)\tilde\psi_1(y) \end{array}\right)
& \mbox{if}~ x<y 
\end{array} \right. , 
\label{eq:gf}
\end{eqnarray}
where $(\psi_1\tilde\psi_2-\psi_2\tilde\psi_1)$ 
is a constant independent of $x$ \cite{kw,oe,kmw}.
We obtain $G^+$ ($G^-$) by solving for wave functions with Im$E$ a small positive 
(negative) number.

The wave functions for the single soliton case, 
$(\phi_1,\phi_2)$ and 
$(\tilde\phi_1,\tilde\phi_2)$,
defined by Eqs. (\ref{eq:solwf1}) and (\ref{eq:solwf2})
do not satisfy the necessary boundary conditions, and therefore cannot be used in 
the representation (\ref{eq:gf}).
However, it is possible to construct 
two new solutions $(\psi_1,\psi_2)$ and $(\tilde\psi_1,\tilde\psi_2)$,
which satisfy the boundary conditions,
by a linear superposition of $(\phi_1,\phi_2)$ and 
$(\tilde\phi_1,\tilde\phi_2)$:
\begin{eqnarray}
\psi_1&=&e^{ikL}\phi_1+e^{-ikL}\tilde\phi_1\nonumber\\
&=&e^{ik(x+L)}\left[\tanh\left(\frac{x}{\xi}\right)-i\frac{E}{\Delta_0}-ik\xi\right]
+ie^{-ik(x+L)}\left[\tanh\left(\frac{x}{\xi}\right)-i\frac{E}{\Delta_0}+ik\xi\right],
\nonumber\\
\psi_2&=&e^{ikL}\phi_2+e^{-ikL}\tilde\phi_2\nonumber\\
&=&-ie^{ik(x+L)}\left[\tanh\left(\frac{x}{\xi}\right)+i\frac{E}{\Delta_0}-ik\xi\right]
+e^{-ik(x+L)}\left[\tanh\left(\frac{x}{\xi}\right)+i\frac{E}{\Delta_0}+ik\xi\right],
\nonumber\\
\tilde\psi_1&=&e^{-ikL}\phi_1+e^{ikL}\tilde\phi_1\nonumber\\
&=&e^{ik(x-L)}\left[\tanh\left(\frac{x}{\xi}\right)-i\frac{E}{\Delta_0}-ik\xi\right]
+ie^{-ik(x-L)}\left[\tanh\left(\frac{x}{\xi}\right)-i\frac{E}{\Delta_0}+ik\xi\right],
\nonumber\\
\tilde\psi_2&=&e^{-ikL}\phi_2+e^{ikL}\tilde\phi_2\nonumber\\
&=&-ie^{ik(x-L)}\left[\tanh\left(\frac{x}{\xi}\right)+i\frac{E}{\Delta_0}-ik\xi\right]
+e^{-ik(x-L)}\left[\tanh\left(\frac{x}{\xi}\right)+i\frac{E}{\Delta_0}+ik\xi\right].
\label{eq:solwf3}
\end{eqnarray}
First, we calculate the Green function for 
$E\ge \Delta_0$.
We introduce a small imaginary part $\tilde E$ to the energy and 
replace $E$ by $E+i\tilde E$.
Then the quantum number $k$ acquires an imaginary part and 
is replaced by $k+i\tilde k$ such that
$E\tilde E=(\hbar v_F)^2k\tilde k$. 
If we choose $\hbar v_F k=-\sqrt{E^2-\Delta_0^2}\le 0$ \cite{sign},
$\tilde E$ and $\tilde k$ have opposite signs. In order to obtain the retarded
Green functions, we assume $\tilde E>0$ 
and take the $L\rightarrow\infty$ limit first
to simplify Eq.~(\ref{eq:solwf3}) 
using the fact that $\tilde k$ is negative and $e^{\tilde k L}$ vanishes in this case.
Finally, we take the $\tilde E\rightarrow 0$ limit to obtain 
the simplified wave functions for use in calculating
the retarded Green functions:
\begin{eqnarray}
\psi_1&=&e^{ik(x+L)}\left[\tanh\left(\frac{x}{\xi}\right)-i\frac{E}{\Delta_0}-ik\xi\right],
\nonumber\\
\psi_2&=&-ie^{ik(x+L)}\left[\tanh\left(\frac{x}{\xi}\right)+i\frac{E}{\Delta_0}-ik\xi\right],
\nonumber\\
\tilde\psi_1&=&ie^{-ik(x-L)}\left[\tanh\left(\frac{x}{\xi}\right)-i\frac{E}{\Delta_0}+ik\xi\right],
\nonumber\\
\tilde\psi_2&=&e^{-ik(x-L)}\left[\tanh\left(\frac{x}{\xi}\right)+i\frac{E}{\Delta_0}+ik\xi\right].
\label{eq:solwf5}
\end{eqnarray}
Similarly, for the advanced Green functions, we assume $\tilde E<0$ and obtain
\begin{eqnarray}
\psi_1&=&ie^{-ik(x+L)}\left[\tanh\left(\frac{x}{\xi}\right)-i\frac{E}{\Delta_0}+ik\xi\right],
\nonumber\\
\psi_2&=&e^{-ik(x+L)}\left[\tanh\left(\frac{x}{\xi}\right)+i\frac{E}{\Delta_0}+ik\xi\right],
\nonumber\\
\tilde\psi_1&=&e^{ik(x-L)}\left[\tanh\left(\frac{x}{\xi}\right)
-i\frac{E}{\Delta_0}-ik\xi\right],
\nonumber\\
\tilde\psi_2&=&-ie^{ik(x-L)}\left[\tanh\left(\frac{x}{\xi}\right)
+i\frac{E}{\Delta_0}-ik\xi\right].
\label{eq:solwf6}
\end{eqnarray}
These wave functions satisfy the required boundary conditions that
$\psi_{1,2}(L)$ and $\tilde\psi_{1,2}(-L)$ diverge as $L$ goes to infinity,
whereas $\psi_{1,2}(-L)$ and $\tilde\psi_{1,2}(L)$ do not diverge.
Substituting Eqs.~(\ref{eq:solwf5}) and (\ref{eq:solwf6}) into Eq.~(\ref{eq:gf}),
we find
\begin{eqnarray}
G^\pm(x,x^\prime)&&=\frac{\mp i}{\hbar v_F}
\frac{\exp\left[\pm i\sqrt{\left(E/\Delta_0
\right)^2-1}~\vert x-x^\prime\vert/\xi\right]}
{4\sqrt{\left(E/\Delta_0\right)^2-1}}
\nonumber\\
&&\times\biggl\{2\left[\frac{E}{\Delta_0}{\bf{1}}
+\sigma_1\pm\sqrt{\left(E/\Delta_0\right)^2-1}~{\rm sgn}
\left(x-x^\prime\right)\sigma_3\right]\nonumber\\
&&~~~
+\frac{\Delta_0}{E}\left[1-t_0t_0^\prime\mp i\sqrt{
\left(E/\Delta_0\right)^2-1}~\vert t_0-t_0^\prime\vert\right]
\left(\sigma_2-{\bf 1}\right)
\nonumber\\
&&~~~    
+\left(t_0+t_0^\prime-2\right)\sigma_1+i
\left(t_0-t_0^\prime\right)\sigma_3\biggr\},
\end{eqnarray}
where ${\bf 1}$ is the unit matrix and $t_0$ and $t_0^\prime$ are defined by
\begin{equation}
t_0=\tanh\left(\frac{x}{\xi}\right),~~~
t_0^\prime=\tanh\left(\frac{x^\prime}{\xi}\right).
\end{equation}
When $E\le -\Delta_0$, we choose $\hbar v_F k=\sqrt{E^2-\Delta_0^2}$ \cite{sign}. Then
$\tilde E$ and $\tilde k$ have opposite signs. 
A calculation similar to that for $E\ge \Delta_0$
leads to the symmetry relationship
\begin{equation}
G_{11}^\pm(x,x^\prime\vert E)=-G_{22}^\mp(x,x^\prime\vert -E),~~~
G_{12}^\pm(x,x^\prime\vert E)=G_{21}^\mp(x,x^\prime\vert -E).
\label{eq:sym}
\end{equation}
The subgap Green function for $0<\vert E\vert<\Delta_0$ will be
derived in Appendix B.

With the exact Green function in hand, it is a straightforward matter to compute
the local electronic density of states (per spin) $\rho_s(x,E)$:
\begin{eqnarray}
\rho_s(x,E)&=&-\frac{1}{\pi}{\rm Im}~ {\rm Tr}~ G^+(x,x\vert E)
\nonumber\\
&=&\left\{\begin{array}{ll}
\displaystyle{\frac{1}{\pi\hbar v_F}\frac{1}{\sqrt{E^2-\Delta_0^2}}
\left[\vert E\vert-\frac{\Delta_0^2}
{2\vert E\vert\cosh^2\left(x/\xi\right)}\right]}
& \mbox{if} ~\vert E\vert\ge \Delta_0\\
0& \mbox{if}~ 0<\vert E\vert <\Delta_0
\end{array}\right..
\label{eq:sdos1}
\end{eqnarray}
The spatially-averaged density of states $\rho_s(E)$ is given by
\begin{eqnarray}
\rho_s(E)&=&\frac{1}{2L}\int_{-L}^L dx~ \rho_s(x,E)
\nonumber\\
&=&\left\{\begin{array}{ll}
\rho_0(E)+\delta\rho_s(E)& \mbox{if} ~\vert E\vert\ge \Delta_0\\
0& \mbox{if}~ 0<\vert E\vert <\Delta_0
\end{array}\right.,
\label{eq:sdos2}
\end{eqnarray}
where the density of states in the Peierls ground state, $\rho_0(E)$, and
the correction to the density of states in the presence of a single soliton, 
$\delta\rho_s(E)$, are 
\begin{eqnarray}
\rho_0(E)&=&\displaystyle{\frac{1}
{\pi\hbar v_F}\frac{\vert E\vert}{\sqrt{E^2-\Delta_0^2}}},
\nonumber\\
\delta\rho_s(E)&=&-\displaystyle{\frac{1}{2\pi\hbar v_F}\frac{\Delta_0^2}
{\vert E\vert\sqrt{E^2-\Delta_0^2}}
\frac{\xi}{L}\tanh\left(\frac{L}{\xi}\right)}.
\label{eq:sdos3}
\end{eqnarray}

In the polaron case, 
we derive the local electronic density of states $\rho_p(x,E)$
for $\vert E\vert\ge \Delta_0$
from the Green functions derived in Appendix A:
\begin{eqnarray}
\rho_p(x,E)&=&\frac{1}{\pi\hbar v_F}\frac{\vert E\vert}{\sqrt{E^2-\Delta_0^2}}
\nonumber\\
&&\times
\left\{1-\frac{\kappa^2\Delta_0^2}
{2\left(E^2-E_0^2\right)}\left[
\frac{1}{\cosh^2\left(\kappa\left(x+x_0\right)/\xi\right)}
+\frac{1}{\cosh^2\left(\kappa\left(x-x_0\right)/\xi\right)}\right]\right\},
\label{eq:pdos1}
\end{eqnarray}
where $E_0=\Delta_0\sqrt{1-\kappa^2}$ and $x_0$ is defined by Eq.~(\ref{eq:pol}).
The density of states when $0\le\vert E\vert<\Delta_0$ and $E\ne\pm E_0$
is zero.
The spatially-averaged density of states, $\rho_p(E)$,
for $\vert E\vert\ge \Delta_0$ is given by
\begin{eqnarray}
\rho_p(E)&=&\rho_0(E)+\delta\rho_p(E),
\nonumber\\
\delta\rho_p(E)&=&
-\frac{1}{\pi\hbar v_F}\frac{\vert E\vert}{\sqrt{E^2-\Delta_0^2}}
\frac{\kappa\Delta_0^2}
{\left(E^2-E_0^2\right)}
\nonumber\\
&&\times
\frac{\xi}{2L}\left[
\tanh\left(\frac{\kappa\left(L+x_0\right)}{\xi}\right)
+\tanh\left(\frac{\kappa\left(L-x_0\right)}{\xi}\right)\right].
\label{eq:pdos2}
\end{eqnarray}

\section{Soliton, polaron and multipolaron energies}
\label{sec-res}

\subsection{Self-consistency in the uniform case}

In this subsection, we reformulate 
the self-consistency equation (\ref{eq:gap}) 
in the uniform case with a constant gap function
in terms of the Green functions.
First, we write the total energy ($E_T$) 
per unit length in terms of the electronic density of states
per spin $\rho(E,\Delta_0)$:
\begin{eqnarray}
\frac{E_T}{2L}&=&\frac{\Delta_i^2}{\pi\hbar v_F \lambda}
+\sum_s\int_{-E_c}^{-\Delta_0} dE~ E\rho(E,\Delta_0)\nonumber\\
&=&\frac{\Delta_i^2}{\pi\hbar v_F \lambda}
-2\int_{\Delta_0}^{E_c} dE~ E\rho(E,\Delta_0),
\label{eq:tote}
\end{eqnarray}
where the cutoff energy $E_c$, introduced to overcome
the well-known difficulty of the continuum model that 
its energy spectrum is unbounded, is 
a function of 
the gap parameter $\Delta_0=\vert \Delta_i+\Delta_e\vert$ 
and satisfies the requirement that
the integrated density of states,
\begin{equation}
N(E_c(\Delta_0),\Delta_0)=
\int_{\Delta_0}^{E_c} dE~\rho(E,\Delta_0),
\end{equation}
is a constant independent of $\Delta_0$. 
As mentioned earlier, in non-degenerate
cases with $\Delta_e>0$, the intrinsic gap $\Delta_i$ 
can be taken to be positive in the ground state
and negative in the metastable state. 
Equivalently, the absolute value $\vert\Delta_i\vert$ satisfies
$\vert\Delta_i\vert=\Delta_0-\Delta_e$ in the ground state
and $\vert\Delta_i\vert=\Delta_0+\Delta_e$ in the metastable state.
Applying the self-consistency condition
\begin{equation}
\frac{\partial}{\partial\Delta_0}
\left[\frac{E_T(\Delta_0)}{2L}\right]=0
\end{equation}
to Eq.~(\ref{eq:tote}) and using the relationships \cite{offdiag}
\begin{eqnarray}
&&\frac{\partial N(E,\Delta_0)}{\partial E}
=\rho(E,\Delta_0),\nonumber\\
&&\frac{d}{d\Delta_0}N(E_c(\Delta_0),\Delta_0)=0,\nonumber\\
&&\frac{\partial N(E,\Delta_0)}{\partial \Delta_0}
=\frac{1}{\pi}{\rm Im}
\left[G_{12}^+(x,x\vert E)+G_{21}^+(x,x\vert E)\right],
\label{eq:idos}
\end{eqnarray}
we derive a new form of the self-consistency 
equation, 
\begin{eqnarray}
\vert\Delta_i\vert&=&-\pi\hbar v_F\lambda\int_{\Delta_0}^{E_c}
dE~\frac{\partial N(E,\Delta_0)}{\partial \Delta_0}
\nonumber\\
&=&-\hbar v_F\lambda\int_{\Delta_0}^{E_c}dE~{\rm Im}
\left[G_{12}^+(x,x\vert E)+G_{21}^+(x,x\vert E)\right],
\end{eqnarray}
in a straightforward manner.
Since in the uniform case
\begin{equation}
G_{12}^+(x,x\vert E)=G_{21}^+(x,x\vert E)=
\frac{-i}{2\hbar v_F}\frac{\Delta_0}{\sqrt{E^2-\Delta_0^2}},
\label{eq:odiag}
\end{equation}
we obtain  
\begin{eqnarray}
\vert\Delta_i\vert&=&\lambda\int_{\Delta_0}^{E_c}
dE~\frac{\Delta_0}{\sqrt{E^2-\Delta_0^2}}\nonumber\\
&=&\lambda \Delta_0\cosh^{-1}\left(\frac{E_c}{\Delta_0}\right)
=\lambda\Delta_0
\ln\left(\frac{E_c+\sqrt{E_c^2-\Delta_0^2}}{\Delta_0}\right),
\label{eq:bcs1}
\end{eqnarray}
which reduces to
\begin{equation}
E_c=\Delta_0\cosh\left(\frac{1}{\lambda}\right)
\label{eq:bcs2}
\end{equation}
in the degenerate case where $\Delta_0=\vert\Delta_i\vert$.
One can easily see that, in the weak coupling limit,
Eq.~(\ref{eq:bcs1}) is approximated by
\begin{equation}
\Delta_0=2E_c\exp\left[-\frac{1}{\lambda}
\left(1\mp\frac{\Delta_e}{\Delta_0}\right)\right],
\end{equation}
where the upper (lower) sign corresponds to the ground (metastable)
state. The total energy per unit length, 
obtained from Eq.~(\ref{eq:tote}), is given by
\begin{equation}
\frac{E_T}{2L}=\frac{\Delta_i^2}{\pi\hbar v_F \lambda}
-\frac{\Delta_0^2}{\pi\hbar v_F}\left\{\cosh^{-1}\left(
\frac{E_c}{\Delta_0}\right)+\frac{1}{2}\sinh\left[2
\cosh^{-1}\left(
\frac{E_c}{\Delta_0}\right)\right]\right\},
\end{equation}
which, using the gap equation (\ref{eq:bcs1}), can be rewritten as
\begin{equation}
\frac{E_T}{2L}=\frac{\Delta_i^2}{\pi\hbar v_F \lambda}
-\frac{\Delta_0^2}{\pi\hbar v_F}\left[
\frac{\vert\Delta_i\vert}{\lambda\Delta_0}+\frac{1}{2}\sinh\left(
\frac{2\vert\Delta_i\vert}{\lambda\Delta_0}\right)\right].
\end{equation}

\subsection{Soliton energy}
\label{sec-se}

We compute the soliton excitation energy $E_s$ 
in the degenerate case 
using the density of states expression obtained 
in Sec.~\ref{sec-gf}. Regardless of the number of electrons
occupying the midgap state with $E=0$, $E_s$ is given by
\begin{eqnarray}
E_s&=&\frac{1}{\pi\hbar v_F\lambda}\int_{-L}^{L} dx~ 
\Delta_0^2\left[\tanh^2(x/\xi)-1\right]
\nonumber\\
&-&4L\left[\int_{\Delta_0}^{E_c^s}dE~E\rho_s(E)
-\int_{\Delta_0}^{E_c^0}dE~E\rho_0(E)\right],
\label{eq:se}
\end{eqnarray}
where $E_c^s$ is the energy cutoff in the presence of a soliton defect, 
while $E_c^0$ ($=\Delta_0\cosh(1/\lambda)$)
is the cutoff in the Peierls ground state. $E_c^s$ and $E_c^0$ are
related by the condition that the total number of electronic states in both cases has
to be the same:
\begin{equation}
\frac{1}{2}+2L\int_{\Delta_0}^{E_c^s}dE~\rho_s(E)=
2L\int_{\Delta_0}^{E_c^0}dE~\rho_0(E),
\label{eq:scut}
\end{equation}
where the number $1/2$ on the left-hand side accounts for the midgap state (per spin).
Substituting Eqs.~(\ref{eq:sdos2}) and (\ref{eq:sdos3}) 
into Eq.~(\ref{eq:scut}) and using the fact that
$E_c^s$ differs from $E_c^0$ by a term of order $1/L$, we derive
\begin{equation}
\frac{E_c^s}{\Delta_0}=\frac{E_c^0}{\Delta_0}
-\frac{\xi}{2L}\tanh\left(\frac{1}{\lambda}\right)
\cot^{-1}\left[\sinh\left(\frac{1}{\lambda}\right)\right]
+O\left(\frac{1}{L^2}\right).
\end{equation}
Using this in Eq.~(\ref{eq:se}), we find, in the $L\rightarrow\infty$ limit,
\begin{equation}
E_s=\frac{2}{\pi}\Delta_0\cosh\left(\frac{1}{\lambda}\right)
\cot^{-1}\left[\sinh\left(\frac{1}{\lambda}\right)\right],
\label{eq:sea}
\end{equation}
which reduces to the well-known value of the soliton excitation energy
in the weak coupling limit, 
$2\Delta_0/\pi$, as $\lambda$ goes to zero and 
increases monotonically to $\Delta_0$
as $\lambda$ grows to infinity. In Fig.~1, we plot the energy 
of a single soliton-anti-soliton pair, $2E_s$,
together with the polaron energy in the degenerate case 
to be obtained in the next subsection (Eq.~(\ref{eq:pdeg})), 
as a function of $\lambda$.

\subsection{Polaron and multipolaron energies}
\label{sec-pe}
The polaron excitation energy $E_p$ is obtained 
in a manner similar to 
that in the previous subsection.
Defining the electron occupation numbers of the subgap states
at $E=E_0$ and $-E_0$ as $n_+$ and $n_-$ 
($n_+,n_-=0,1,2$), we have
\begin{eqnarray}
E_p&=&\frac{1}{\pi\hbar v_F\lambda}\int_{-L}^{L} dx~
\Delta_{i}^2\left\{\left[1-\frac{\Delta_0}{\Delta_i}
\kappa\left(t_+-t_-\right)
\right]^2-1\right\}\nonumber\\
&+&(n_+-n_-)E_0
-4L\left[\int_{\Delta_0}^{E_c^p}dE~E\rho_p(E)
-\int_{\Delta_0}^{E_c^0}dE~E\rho_0(E)\right],
\label{eq:pe}
\end{eqnarray}
where the constant $\Delta_i$ ($=\Delta_0-\Delta_e>0$) 
satisfying Eq.~(\ref{eq:bcs1})
is the intrinsic gap in the ground state and 
$E_c^p$ is the energy cutoff 
in the presence of a polaron defect.
We emphasize that the coherence length $\xi$ appearing in the definitions
of $t_+$ and $t_-$ (Eq.~(\ref{defs})) 
is defined by $\hbar v_F/\Delta_0$, not by 
$\hbar v_F/\Delta_i$.
The equation (\ref{eq:scut}) is replaced by
\begin{equation}
1+2L\int_{\Delta_0}^{E_c^p}dE~\rho_p(E)=
2L\int_{\Delta_0}^{E_c^0}dE~\rho_0(E).
\label{eq:pcut}
\end{equation}
Substituting Eq.~(\ref{eq:pdos2}) 
into the above equation, we find
\begin{equation}
\frac{E_c^p}{\Delta_0}=\frac{E_c^0}{\Delta_0}
-\frac{\xi}{L}\tanh\left(\frac{\Delta_i}
{\lambda\Delta_0}\right)
\cot^{-1}\left[\frac{1}{\kappa}
\sinh\left(\frac{\Delta_i}
{\lambda\Delta_0}\right)\right]
+O\left(\frac{1}{L^2}\right).
\end{equation}
The polaron energy $E_p$ to order 1 follows from this 
and Eq.~(\ref{eq:pe}) and is equal to
\begin{eqnarray}
E_p&=&(n_+-n_-+2)E_0+\frac{4}{\pi}\Delta_0\cosh\left(
\frac{\Delta_i}{\lambda\Delta_0}\right)\cot^{-1}
\left[\frac{1}{\kappa}\sinh\left(
\frac{\Delta_i}{\lambda\Delta_0}\right)\right]
\nonumber\\
&-&\frac{4}{\pi}E_0\tan^{-1}
\left[\frac{\kappa\Delta_0}{E_0}\coth\left(
\frac{\Delta_i}{\lambda\Delta_0}\right)\right]
+\frac{4}{\pi}\Delta_0\Gamma\left(\tanh^{-1}\kappa
-\kappa\right),
\label{eq:pee}
\end{eqnarray}
where $\Gamma\equiv\Delta_e/\lambda\Delta_0$.
The self-consistency condition is equivalent to the condition
that $E_p$ has to be minimized with respect to $\kappa$
for given values of $n_+$ and $n_-$.
Applying a weaker condition that $E_p$ takes an extremal value,
that is, $\partial E_p/\partial\kappa=0$
to Eq.~(\ref{eq:pee}), we obtain
\begin{equation}
\tan^{-1}\left[\tan\theta~\coth\left(\frac{1}
{\lambda}-\Gamma\right)\right]+\Gamma\tan\theta
=\frac{\pi}{4}\left(n_+-n_-+2\right),~~
0\le\theta\le\frac{\pi}{2},
\label{eq:pscc}
\end{equation}
where $\theta$ is defined by $\kappa=\sin\theta$
and $E_0=\Delta_0\cos\theta$.
It turns out that, in the degenerate case with $\Gamma=0$,
stable solutions which satisfy Eq.~(\ref{eq:pscc})
and minimize Eq.~(\ref{eq:pee}) exist only for the electron polaron
state ($n_+=1$, $n_-=2$), with the effective charge $Q\equiv 2-n_+-n_-=-1$
and the spin $S=1/2$,
and the hole polaron state ($n_+=0$, $n_-=1$),
with $Q=+1$ and $S=1/2$. In both cases, $\theta=\tan^{-1}[\tanh(1/\lambda)]$
and
\begin{eqnarray}
&&\kappa=\frac{\tanh(1/\lambda)}{\sqrt{\tanh^2(1/\lambda)+1}},~~
E_0=\frac{\Delta_0}{\sqrt{\tanh^2(1/\lambda)+1}},
\nonumber\\
&&E_p=\frac{4}{\pi}\Delta_0\cosh(1/\lambda)
\cot^{-1}\left[\sqrt{\cosh^2(1/\lambda)
+\sinh^2(1/\lambda)}\right],
\label{eq:pdeg}
\end{eqnarray}
which reduce to $\kappa=1/\sqrt{2}$, $E_0=\Delta_0/\sqrt{2}$ and
$E_p=2\sqrt{2}\Delta_0/\pi$ in the weak coupling limit.
The behavior of $E_p$  
as a function of $\lambda$ in the degenerate case is shown
in Fig.~1.

In the non-degenerate case with $\Gamma>0$, stable polaron 
and multipolaron (bipolaron, tripolaron, quadripolaron) 
solutions exist
for all possible combinations of $n_+$ and $n_-$ except ($n_+=0$, $n_-=2$).
There are three bipolaron solutions with the effective occupation
number $N\equiv n_+-n_-+2=2$. Both the electron-electron bipolaron 
($n_+=n_-=2$, $Q=-2$)
and the hole-hole bipolaron ($n_+=n_-=0$, $Q=+2$) have spin zero, while the 
electron-hole bipolaron ($n_+=n_-=1$, $Q=0$), which can also be called
an exciton, exists in either spin singlet ($S=0$) or triplet ($S=1$) forms.
More details about the interpretation of multipolaron solutions can be 
found in Refs.~\cite{roth,camp}. 
Explicit values of $\kappa$, $E_0$ and $E_p$ 
for given values of $\lambda$ and $\Gamma$ can be obtained only 
numerically.
As a specific example of applying Eqs.~(\ref{eq:pee}) and (\ref{eq:pscc}),
we consider the case of $cis$-polyacetylene. There 
is a considerable uncertainty
in the precise values of $\lambda$ and $\Gamma$ ($=\Delta_e/\lambda\Delta_0$). 
In order to make a rough 
estimate of the excitation energy, 
we use a set of the approximate values close to those assumed in \cite{boya}, which are 
$\lambda=0.4$, $\Delta_e=0.06$ eV, $\Delta_0=1$ eV, and therefore,
$\Gamma=0.15$.
The results are listed in Table I.

\section{Conclusion}
\label{sec-conc}

In the present paper, we have reformulated the theory of
solitons, polarons and multipolarons in both degenerate and 
non-degenerate conducting polymers.
Especially, we have developed a simple method for calculating
the Green function and the density of states in the presence of
a soliton or polaron defect and computed the soliton, polaron
and multipolaron excitation energies and the self-consistent 
gap functions exactly for an arbitrary value of the electron-phonon
coupling constant (See Eqs.~(\ref{eq:sea},\ref{eq:pee}-\ref{eq:pdeg})).
Our method can be generalized 
in a straightforward way to more complicated situations.
In a previous work, Kim and Wilkins 
have devised an efficient numerical
method for calculating the {\it exact} disorder-averaged Green function 
in the presence of a soliton or polaron defect and disorder \cite{kw}. 
It appears that, by combining this 
method with the results of the present work, one can 
study the disordered soliton or polaron problem in an exact and
self-consistent manner.
Work in this direction will be presented in a separate 
publication.

\acknowledgments
We are grateful to John Wilkins for reading the manuscript
and making valuable comments. 
This work has been supported by  
the Korea Science and Engineering Foundation
through Grant No. 961-0209-052-2
and the U.S. DOE, Basic Energy Sciences,
Division of Materials Sciences. 

\appendix
\section{Polaron Green function}
The spatial variation of the order parameter for a single 
polaron located at $x=0$ is given by 
\begin{eqnarray}
\frac{\Delta(x)}{\Delta_0}&=& 1-\kappa \left\{ \tanh \left[
 \frac{\kappa(x+x_0)}
{\xi} \right]-\tanh \left[ \frac{\kappa(x-x_0)}{\xi}
 \right] \right\},\nonumber \\
\frac{x_0}{\xi}&=&\frac{1}{4\kappa}\ln\left(\frac{1+\kappa}{1-\kappa}\right),
~~~0<\kappa<1.
\label{eq:pol}
\end{eqnarray}
The dimensionless parameter $\kappa$ appearing in this equation
determines the shape of the polaron, which changes from
an extremely shallow well centered at $x=0$ for $0<\kappa\ll 1$ 
to a well-separated pair
of soliton and anti-soliton located at $x_0$ and $-x_0$, respectively,
as $\kappa\rightarrow 1$. 
This parameter cannot take an arbitrary value, 
since the self-consistent
gap equation is not satisfied
for every value of $\kappa$.
When $\Delta_e=0$ and in the weak electron-phonon coupling limit
(that is, the $\lambda\rightarrow 0$ limit), it turns
out that $\kappa$ has the unique value of $1/\sqrt{2}$, as
will be shown in Sec.~\ref{sec-pe}.

In case of the single polaron configuration, the exact unnormalized 
wave functions for unbound states with $\vert E\vert\ge\Delta_0$
can be written as \cite{cb2} 
\begin{eqnarray}
\psi_{1s}(x)&=&\phi_1(x)=e^{ikx}\left[k\xi+\frac{E}{\Delta_0}-1+
\gamma(1+i)t_+ -\delta(1-i)t_- \right],\nonumber\\ 
\psi_{2s}(x)&=&\phi_2(x)=e^{ikx}\left[k\xi-\frac{E}{\Delta_0}+1-
\gamma(1-i)t_+ +\delta(1+i)t_- \right],
\label{eq:polwf1}
\end{eqnarray}
where $t_+$, $t_-$, $\gamma$ and $\delta$ are defined by
\begin{eqnarray}
&&t_+=\tanh\left[\frac{\kappa(x+x_0)}{\xi}\right],~~~
t_-=\tanh\left[\frac{\kappa(x-x_0)}{\xi}\right],\nonumber\\
&&\gamma=\frac{\kappa}{2}\left(1+\frac{i\hbar k_F k}{E+\Delta_0}\right),~~~
\delta=\frac{\kappa}{2}\left(1-\frac{i\hbar k_F k}{E+\Delta_0}\right).
\label{defs}
\end{eqnarray}
Similarly to the soliton case, we introduce an equivalent set of
wave functions
\begin{eqnarray}
\psi_{1s}(x)&=&\tilde\phi_1(x)=e^{-ikx}\left[k\xi-\frac{E}{\Delta_0}+1-
\delta(1+i)t_+ +\gamma(1-i)t_- \right],\nonumber\\ 
\psi_{2s}(x)&=&\tilde\phi_2(x)=e^{-ikx}\left[k\xi+\frac{E}{\Delta_0}-1+
\delta(1-i)t_+ -\gamma(1+i)t_- \right],
\label{eq:polwf2}
\end{eqnarray}
which are obtained from Eq.~(\ref{eq:polwf1}) by replacing $k$ with $-k$ and
multiplying the wave functions by $-1$.
In addition, there are two subgap polaron bound states 
located at $E=\pm E_0\equiv\pm \Delta_0\sqrt{1-\kappa^2}$,
the wave functions for which we do not write down here explicitly.
As in the soliton case, we can also use the wave functions (\ref{eq:polwf1})
and (\ref{eq:polwf2}) for computing the subgap Green functions when
$0\le\vert E\vert<\Delta_0$ and $E\ne \pm E_0$.

We superpose two independent solutions
in the polaron case,  
(\ref{eq:polwf1}) and (\ref{eq:polwf2}), 
to make two new wave functions $(\psi_1,\psi_2)$ and $(\tilde\psi_1,\tilde\psi_2)$
satisfying the necessary boundary conditions.
When $E\ge\Delta_0$, 
following the same procedure as in Sec.~\ref{sec-gf}, we obtain
the simplified wave functions
\begin{eqnarray}
\psi_1&=&e^{ik(x+L)}\left[k\xi+\frac{E}{\Delta_0}-1+
\gamma(1+i)t_+ -\delta(1-i)t_- \right],\nonumber\\ 
\psi_2&=&e^{ik(x+L)}\left[k\xi-\frac{E}{\Delta_0}+1-
\gamma(1-i)t_+ +\delta(1+i)t_- \right],\nonumber\\
\tilde\psi_1&=&e^{-ik(x-L)}\left[k\xi-\frac{E}{\Delta_0}+1-
\delta(1+i)t_+ +\gamma(1-i)t_- \right],\nonumber\\ 
\tilde\psi_2&=&e^{-ik(x-L)}\left[k\xi+\frac{E}{\Delta_0}-1+
\delta(1-i)t_+ -\gamma(1+i)t_- \right],
\label{polg1}
\end{eqnarray}
with $\hbar v_F k=-\sqrt{E^2-\Delta_0^2}$ \cite{sign}
for computing the retarded Green functions and
\begin{eqnarray}
\psi_1&=&e^{-ik(x+L)}\left[k\xi-\frac{E}{\Delta_0}+1-
\delta(1+i)t_+ +\gamma(1-i)t_- \right],\nonumber\\ 
\psi_2&=&e^{-ik(x+L)}\left[k\xi+\frac{E}{\Delta_0}-1+
\delta(1-i)t_+ -\gamma(1+i)t_- \right],\nonumber\\
\tilde\psi_1&=&e^{ik(x-L)}\left[k\xi+\frac{E}{\Delta_0}-1+
\gamma(1+i)t_+ -\delta(1-i)t_- \right],\nonumber\\ 
\tilde\psi_2&=&e^{ik(x-L)}\left[k\xi-\frac{E}{\Delta_0}+1-
\gamma(1-i)t_+ +\delta(1+i)t_- \right],
\label{polg2}
\end{eqnarray}
for the advanced Green functions. The Green functions 
obtained by substituting Eqs.~(\ref{polg1}) and (\ref{polg2})
into Eq.~(\ref{eq:gf}) are
\begin{eqnarray}
G^\pm(x,x^\prime)&&=\frac{\mp i}{\hbar v_F}
\frac{\exp\left[\pm i\sqrt{\left(E/\Delta_0
\right)^2-1}~\vert x-x^\prime\vert/\xi\right]}
{4\sqrt{\left(E/\Delta_0\right)^2-1}}
\nonumber\\
&&\times\Biggl\{ 
2\left(\frac{E}{\Delta_0}{\bf{1}}+\sigma_1
\pm\sqrt{\left(E/\Delta_0\right)^2-1}~{\rm sgn}
\left(x-x^\prime\right)\sigma_3\right)
\nonumber\\
&&~~~+\frac{\kappa^2\Delta_0^2}
{E_0^2-E^2}\Biggl\lbrack
\frac{E}{\Delta_0}\Biggl\lgroup
\left(1-t_-t_-^\prime\mp i\sqrt{\left(E/\Delta_0\right)^2-1}~
\vert t_- -t_-^\prime\vert/\kappa
\right)\left({\bf 1}-\sigma_2\right)
\nonumber\\
&&~~~+\left(1-t_+t_+^\prime
\mp i\sqrt{\left(E/\Delta_0\right)^2-1}~
\vert t_+ -t_+^\prime\vert/\kappa
\right)\left({\bf 1}+\sigma_2\right)\Biggr\rgroup
\nonumber\\
&&~~~+\Biggl\lgroup\left
(1-\left(E/\Delta_0\right)^2\right)\left(t_-+
t_-^\prime-t_+-t_+^\prime\right)/\kappa
+2-t_-t_+^\prime-t_+t_-^\prime
\nonumber\\
&&~~~
\mp i\sqrt{\left(E/\Delta_0\right)^2-1}
\bigg(t_+t_-^\prime
-t_-t_+^\prime
+\left(t_--t_-^\prime+t_+-t_+^\prime\right)/\kappa\bigg){\rm sgn}
\left(x-x^\prime\right)
\Biggr\rgroup\sigma_1
\nonumber\\
&&~~~+i\Biggl\lgroup
\mp i\sqrt{\left(E/\Delta_0\right)^2-1}\bigg(2-t_-t_+^\prime
-t_+t_-^\prime
+\left(t_-+t_-^\prime-t_+-t_+^\prime\right)/\kappa\bigg)
{\rm sgn}\left(x-x^\prime\right)
\nonumber\\
&&~~~+\left(1-\left(E/\Delta_0\right)^2\right)
\left(t_--t_-^\prime+t_+-t_+^\prime\right)/\kappa
+t_+t_-^\prime-t_-t_+^\prime
\Biggr\rgroup\sigma_3\Biggr\rbrack\Biggr\},
\end{eqnarray} 
The Green functions for $E\le -\Delta_0$ are obtained 
using the symmetry relationship
Eq.~(\ref{eq:sym}), which is valid in the polaron
case, too.  

When $0\le\vert E\vert<\Delta_0$ and $E\ne \pm E_0$, 
we obtain 
\begin{eqnarray}
\psi_1&=&e^{q(x+L)}\left[-iq\xi+\frac{E}{\Delta_0}-1+
\gamma(1+i)t_+ -\delta(1-i)t_- \right],\nonumber\\ 
\psi_2&=&e^{q(x+L)}\left[-iq\xi-\frac{E}{\Delta_0}+1-
\gamma(1-i)t_+ +\delta(1+i)t_- \right],\nonumber\\
\tilde\psi_1&=&e^{-q(x-L)}\left[-iq\xi-\frac{E}{\Delta_0}+1-
\delta(1+i)t_+ +\gamma(1-i)t_- \right],\nonumber\\ 
\tilde\psi_2&=&e^{-q(x-L)}\left[-iq\xi+\frac{E}{\Delta_0}-1+
\delta(1-i)t_+ -\gamma(1+i)t_- \right],
\label{eq:polwf4}
\end{eqnarray}
where $q=ik=\sqrt{1-\left(E/\Delta_0\right)^2}/\xi$ (See Appendix B) and 
$\gamma$ and $\delta$ are given by Eq.~(\ref{defs}) with
$ik$ replaced by $q$.
Substituting Eq.~(\ref{eq:polwf4}) into Eq.~(\ref{eq:gf}),
we obtain
\begin{eqnarray}
G^\pm(x,x^\prime)&&=-\frac{1}{\hbar v_F}
\frac{\exp\left[-\sqrt{1-\left(E/\Delta_0
\right)^2}\vert x-x^\prime\vert/\xi\right]}
{4\sqrt{1-\left(E/\Delta_0\right)^2}}
\nonumber\\
&&\times\Biggl\{ 
2\left(\frac{E}{\Delta_0}{\bf{1}}+\sigma_1
+i\sqrt{1-\left(E/\Delta_0\right)^2}{\rm sgn}
\left(x-x^\prime\right)\sigma_3\right)
\nonumber\\
&&~~~+\frac{\kappa^2\Delta_0^2}
{E_0^2-E^2}\Biggl\lbrack
\frac{E}{\Delta_0}\Biggl\lgroup
\left(1-t_-t_-^\prime+\sqrt{1-\left(E/\Delta_0\right)^2}
\vert t_- -t_-^\prime\vert/\kappa
\right)\left({\bf 1}-\sigma_2\right)
\nonumber\\
&&~~~+\left(1-t_+t_+^\prime
+\sqrt{1-\left(E/\Delta_0\right)^2}
\vert t_+ -t_+^\prime\vert/\kappa
\right)\left({\bf 1}+\sigma_2\right)\Biggr\rgroup
\nonumber\\
&&~~~+\Biggl\lgroup\left
(1-\left(E/\Delta_0\right)^2\right)\left(t_-+
t_-^\prime-t_+-t_+^\prime\right)/\kappa
+2-t_-t_+^\prime-t_+t_-^\prime
\nonumber\\
&&~~~
+\sqrt{1-\left(E/\Delta_0\right)^2}\bigg(t_+t_-^\prime
-t_-t_+^\prime+\left(t_--t_-^\prime+t_+-t_+^\prime\right)/\kappa\bigg){\rm sgn}
\left(x-x^\prime\right)
\Biggr\rgroup\sigma_1
\nonumber\\
&&~~~+i\Biggl\lgroup
\sqrt{1-\left(E/\Delta_0\right)^2}\bigg(2-t_-t_+^\prime
-t_+t_-^\prime
+\left(t_-+t_-^\prime-t_+-t_+^\prime\right)/\kappa\bigg)
{\rm sgn}\left(x-x^\prime\right)
\nonumber\\
&&~~~+\left(1-\left(E/\Delta_0\right)^2\right)
\left(t_--t_-^\prime+t_+-t_+^\prime\right)/\kappa
+t_+t_-^\prime-t_-t_+^\prime
\Biggr\rgroup\sigma_3\Biggr\rbrack\Biggr\},
\end{eqnarray}
where $t_+^\prime$ and $t_-^\prime$ are defined similarly to $t_+$
and $t_-$ except that $x$ is replaced by $x^\prime$.

\section{Subgap Green function in the soliton case}
Here we calculate the Green function for $0<\vert E\vert<\Delta_0$.
In this case, we have $k\xi=\pm i\sqrt{1-(E/\Delta_0)^2}\equiv \pm i q\xi$.
We choose $k=-iq$ \cite{sign} 
and take the $L\rightarrow\infty$ limit to simplify 
Eq.~(\ref{eq:solwf3}) using the fact that $q$ is positive and $e^{-qL}$ vanishes:
\begin{eqnarray}
\psi_1&=&e^{q(x+L)}\left[\tanh\left(\frac{x}{\xi}\right)-i\frac{E}{\Delta_0}-q\xi\right],
\nonumber\\
\psi_2&=&-ie^{q(x+L)}\left[\tanh\left(\frac{x}{\xi}\right)+i\frac{E}{\Delta_0}-q\xi\right],
\nonumber\\
\tilde\psi_1&=&ie^{-q(x-L)}\left[\tanh\left(\frac{x}{\xi}\right)-i\frac{E}{\Delta_0}+q\xi\right],
\nonumber\\
\tilde\psi_2&=&e^{-q(x-L)}\left[\tanh\left(\frac{x}{\xi}\right)+i\frac{E}{\Delta_0}+q\xi\right].
\label{eq:solwf4}
\end{eqnarray}
The Green functions are obtained in a straightforward manner 
by substituting Eq.~(\ref{eq:solwf4}) directly into
Eq.~(\ref{eq:gf}). We find that the retarded and advanced Green functions are 
the same in the present case and given by 
\begin{eqnarray}
G^\pm(x,x^\prime)&&=-\frac{1}{\hbar v_F}\frac{\exp\left[-\sqrt{1-\left(E/\Delta_0
\right)^2}\vert x-x^\prime\vert/\xi\right]}
{4\sqrt{1-\left(E/\Delta_0\right)^2}}
\nonumber\\
&&\times\biggl\{2\left[\frac{E}{\Delta_0}{\bf{1}}
+\sigma_1+i\sqrt{1-\left(E/\Delta_0\right)^2}{\rm sgn}
\left(x-x^\prime\right)\sigma_3\right]\nonumber\\
&&~~~
+\frac{\Delta_0}{E}\left[1-t_0t_0^\prime+\sqrt{1-
\left(E/\Delta_0\right)^2}\vert t_0-t_0^\prime\vert\right]
\left(\sigma_2-{\bf 1}\right)
\nonumber\\
&&~~~    
+\left(t_0+t_0^\prime-2\right)\sigma_1+i
\left(t_0-t_0^\prime\right)\sigma_3\biggr\}.
\end{eqnarray}

\begin{figure}
\protect\centerline{\epsfxsize=4in \epsfbox{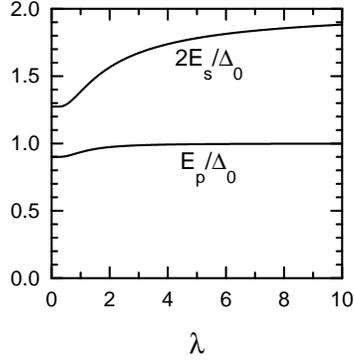}}
\caption{The energy of a soliton-anti-soliton pair, $2E_s$ and the polaron
energy in the degenerate case, $E_p$ vs the electron-phonon coupling constant, $\lambda$.}
\end{figure}

\begin{table}
\caption{Polaron and multipolaron states in $cis$-polyacetylene
when $\lambda=0.4$ and $\Gamma=0.15$. $N=n_+-n_-+2$ is
the effective occupation number, $Q=2-n_+-n_-$ is the effective
charge, $\theta$ is the angle (in radians) defined by Eq.~(\ref{eq:pscc}) and
$E_p$ is the excitation energy.}
\begin{tabular}{ccccccc}
$n_+$ & $n_-$ & N & Q & $\theta$ & $E_p/\Delta_0$ 
& Interpretation\\ 
\tableline
1     & 2  & 1 & $-1$   & 0.66    & 0.93 & electron (e) polaron      \\
0     & 1  & 1 & $+1$   & 0.66    & 0.93 & hole (h) polaron          \\
2     & 2  & 2 & $-2$   & 1.19    & 1.51 & e-e bipolaron             \\
1     & 1  & 2 & $ 0$   & 1.19    & 1.51 & e-h bipolaron (or exciton)\\
0     & 0  & 2 & $+2$   & 1.19    & 1.51 & h-h bipolaron             \\
2     & 1  & 3 & $-1$   & 1.41    & 1.75 & e-e-h tripolaron          \\
1     & 0  & 3 & $+1$   & 1.41    & 1.75 & e-h-h tripolaron          \\ 
2     & 0  & 4 & $ 0$   & 1.48    & 1.87 & e-e-h-h quadripolaron     \\ 
\end{tabular}
\label{table1}
\end{table}


\begin{references}
\bibitem{int}
It is well-known that the electron-electron interaction is
relevant in polyacetylene in several respects. For detailed discussions,
see, e.g., References 10 and 11.
\bibitem{cont} 
We caution that the continuum model is strictly valid 
only when the coherence length is much larger than the lattice
constant. Since this condition is not met when $\lambda\gg 1$, 
we do not expect our results to be quantitatively applicable
to real conducting polymers in that case.
\bibitem{sign} The same expression for the Green function is obtained
for a different sign choice for $k$.
\bibitem{offdiag}
Eq.~(\ref{eq:idos}) follows from the definition of
$N(E,\Delta_0)$, which is equal to $\sqrt{E^2-\Delta_0^2}/\pi\hbar v_F$
for $E>\Delta_0$, and Eq.~(\ref{eq:odiag}).
\end{references}
\end{document}